\documentclass[preprint,floatfix,aps,showpacs]{revtex4}
\newcommand {\bea}{\begin{eqnarray}}
\newcommand {\eea}{\end{eqnarray}}
\newcommand {\be}{\begin{equation}}
\newcommand {\ee}{\end{equation}}

\usepackage{graphicx}
\usepackage{epsfig,subfigure}
\usepackage{amsmath}
\begin{document}
\def\({\left(}
\def\){\right)}
\def\[{\left[}
\def\]{\right]}

\def\Journal#1#2#3#4{{#1} {\bf #2}, #3 (#4)}
\def\RPP{{Rep. Prog. Phys}}
\def\PRC{{Phys. Rev. C}}
\def\PRD{{Phys. Rev. D}}
\def\FP{{Foundations of Physics}}
\def\ZPA{{Z. Phys. A}}
\def\NPA{{Nucl. Phys. A}}
\def\JPG{{J. Phys. G Nucl. Part}}
\def\PRL{{Phys. Rev. Lett.}}
\def\PRpt{{Phys. Rep.}}
\def\PLB{{Phys. Lett. B}}
\def\AP{{Ann. Phys (N.Y.)}}
\def\EPJA{{Eur. Phys. J. A}}
\def\NP{{Nucl. Phys}}
\def\ZP{{Z. Phys}}
\def\RMP{{Rev. Mod. Phys}}
\def\IJMPE{{Int. J. Mod. Phys. E}}
\input epsf

\title{Low-Density Instability of Multi-Component Matter with Trapped Neutrinos}

\author{T. Mart and A. Sulaksono }

\affiliation{Departemen Fisika, FMIPA, Universitas Indonesia,
Depok, 16424, Indonesia}

\begin{abstract}
The effect of neutrino trapping on the longitudinal dielectric 
function at low densities has been investigated 
by using different relativistic mean field models. Parameter
sets G2 of Furnstahl-Serot-Tang and Z271 of Horowitz-Piekarewicz, along with 
the adjusted parameter sets of both models, have been used in this study.
The role of the isovector adjustment and the effect of the Coulomb interaction
have been also studied. The effect of the isovector adjustment is found 
to be more significant in the Horowitz-Piekarewicz model, 
not only in the neutrinoless matter, but also
in the matter with neutrino trapping. Although almost independent to
the variation of 
the leptonic fraction, the instability region of matter with neutrino 
trapping is found to be larger. The presence of more protons and
electrons compared to the neutrinoless case is the reason
behind this finding. For parameter sets with soft equation of states at
low density, the appearance of a large and negative $\varepsilon_L (q,q_0=0)$ in 
some parts of the edge of the instability region in matter with neutrino 
trapping is understood
as a consequence of the fact that the Coulomb interaction produced
by electrons and protons interaction is larger than the repulsive
isovector interaction created by the asymmetry between the proton and neutron
numbers.

\end{abstract}
\maketitle
\newpage
%%%%%%%%%%%%%%%%%%%%%%%%%%%%%%%%%%%%%%%%%%%%%%%%%%%%%%%%%%%%%%%%%%%%%%%%%
\section{Introduction}
\label{sec_intro}
%%%%%%%%%%%%%%%%%%%%%%%%%%%%%%%%%%%%%%%%%%%%%%%%%%%%%%%%%%%%%%%%%%%%%%%%%
At low densities both the relativistic and the non-relativistic mean
field models predict a liquid-gas phase transition region for nuclear
matter leading, for dense star matter, to a non-homogeneous phase
commonly called pasta phase, which is formed by a competition between the long
range Coulomb repulsion and the short range nuclear
attraction~\cite{Provi07}. This transition has substantial consequences on the
properties of stellar matter and neutrino
transport~\cite{Duco07}. Considerable efforts to comprehend the uniform ground state
stability of multi-component systems consisting of electrons,
neutrinos, protons, and neutrons as a good approximation of this 
transition have been recently devoted, not only in the zero temperature approximation, 
but also for finite 
temperature~\cite{Pethick,Douchin,Carr,Horowitz01,Provi1,Provi2,Provi3,Provi4,anto06,Muller:1995ji}.
It is obvious that in order to understand the physics
inside the non-homogeneous (unstable) regions like the mechanism of nuclear
creation with slab-like or rod-like shape, we have to go beyond the mean
field approximation. Attempts in this direction are discussed
in Refs.~\cite{Provi07,Napoli07,Watanabe04,Horowitz04,Sonoda}. Moreover, 
in the collapsing supernova core and at sub-nuclear densities, the
transition of nuclear shape from sphere to other exotic shapes has
significant effects to the neutrino mean free path. However, how these effects 
modify the neutrino mean free path is not fully understood yet~\cite{Horowitz04,Sonoda}.

Another motivation to study this transition comes from the fact that a neutron star 
is expected to have a solid inner crust of nonuniform neutron-rich matter above 
its liquid mantle~\cite{Carr} and the mass of its crust depends sensitively on 
the density of its inner edge and on its equation of state (EOS)~\cite{Douchin}. 
On the other hand, the critical density ($\rho_c$), a density at which the uniform 
liquid becomes unstable to a small density fluctuation, can be used as a good 
approximation of the edge density of the crust~\cite{Carr}. By generalizing 
the dynamical stability analysis of Ref.~\cite{Horo1} in order to accommodate 
the various nonlinear terms in the relativistic mean field (RMF) model of 
Horowitz-Piekarewicz~\cite{Horowitz01}, Carriere {\it et al.}~\cite{Carr} found a strong 
correlation between $\rho_c$ in the neutron star and the density dependence 
of nuclear matter symmetry energy ($a_{\rm sym}$). This leads to a suggestion
that a measurement of the neutron radius in $^{208}{\rm Pb}$ will provide useful 
information on the $\rho_c$ \cite{Carr,Horowitz01}.

In our previous work~\cite{anto06}, the critical densities of uniform 
matter with and without neutrino trapping have been calculated and analyzed
by means of different RMF models. In this analysis it is shown that
the interplay between the
dominant contribution of the matter composition and the
effective masses of mesons and nucleons leads to higher critical
densities for matter with neutrino trapping. Furthermore, it was also found that the
predicted critical density is insensitive to the number of
trapped neutrinos as well as to the RMF model used. However, the discussion
about the reason behind these findings was not quite robust. On the other
hand, as we mentioned above, the neutrino transport is very 
crucial in the dynamics of the core-collapsing supernova 
due to the fact that the neutrinos carry most of the energy
away and will lose also their energies by exciting collective nuclear
and plasmon modes~\cite{Provi2}. Similar situation can be also found 
in the neutron stars. Moreover, it was also shown in Ref.~\cite{Provi2}
that the behavior of electrons in matter depends strongly on the wave
length or momentum of the external perturbation $q$. Note that this
momentum is related to the energy transfer of the neutrinos that propagate in
matter. 

The present paper reports on the extension of our previous
investigation~\cite{anto06} by calculating the longitudinal dielectric
function of ERMF models and analyzing the relation between the
obtained results and the isovector sector adjustment, the 
presence of the long-range Coulomb interaction, as well as the presence of 
electrons. The purpose of this work is to explain the reason behind
the appearance of each point along the onset of the instability. To this 
end, we should emphasize here that we need to calculate the dielectric function
in the edge of non-homogeneous regions because information from the 
critical density alone is insufficient.
Furthermore, it is also important to emphasize that our definition
of the instability is not the non-homogeneous area, but 
rather it is connected with the points where these non-homogeneities start
to appear. As a consequence, the assumption of the uniform matter 
in the calculation is still valid. 

This paper is organized as follows. The RMF models and some constraints
used in the present analysis are briefly discussed in
Sec.~\ref{sec_models}.  In Sec.~\ref{sec_LDF}, a discussion
of the longitudinal dielectric function is given.  
In Sec.~\ref{sec_results} we present the graphical results 
of the onset of instability along with the 
corresponding discussions. Finally, we give the conclusion in
Sec.~\ref{sec_conclu}.

%%%%%%%%%%%%%%%%%%%%%%%%%%%%%%%%%%%%%%%%%%%%%%%%%%%%%%%%%%%%%%%%%%%%%%%%%
\section{RMF Models}
\label{sec_models}
%%%%%%%%%%%%%%%%%%%%%%%%%%%%%%%%%%%%%%%%%%%%%%%%%%%%%%%%%%%%%%%%%%%%%%%%%
To describe the multi-component matter, we use the  Lagrangian density~\cite{anto06} 
\bea
{\mathcal L} = {\mathcal L}_N + {\mathcal L}_M + {\mathcal L}_{\rm HP}+ {\mathcal L}_{L}~,
\label{eq:nuclag}
\eea
where the first three terms describe the nucleons in the mean field level, 
while the last term indicates a free Lagrangian for leptons. The first term 
is the Lagrangian for nucleons interacting with each other via meson exchanges. 
The second term is the Lagrangian for mesons, containing also their nonlinear 
self coupling information. The third term is added to accommodate the  Horowitz-Piekarewicz
isovector nonlinear term~\cite{Horowitz01}. 

In this study we investigate
two RMF models, namely, the G2 parameter set of 
Furnstahl-Serot-Tang~\cite{Furn96} (also known as the ERMF model) 
and the Z271 parameter set of
Horowitz-Piekarewicz~\cite{Horowitz01}. The effective Lagrangian density of
Furnstahl-Serot-Tang model has been constructed to fulfill the 
symmetries of quantum chromodynamics and is expanded in powers of
the fields and their derivatives up to order $\nu= 4$. Furthermore,
the Lagrangian exploits the natural coupling constants and its 
application to study the properties of finite nuclei is quite
successful. The inclusion of the higher
order terms was found to be unimportant (and undetermined) for the
nuclear observables of interest~\cite{Furn96}. On the other
hand, the 
Horowitz-Piekarewicz model is an extension of the
standard RMF model with an additional isovector-vector nonlinear
term. 

Therefore, the two models can be considered as the generalization 
of the standard RMF models. Details of
the individual terms and coupling constants of both models can be found 
in Ref.~\cite{anto06}. For each model we use two different parameter 
sets, i.e., the standard ones (G2 and Z271) and the adjusted ones 
which produce softer symmetry energy predictions at high densities
(G2* and Z271*). 
%----------------------------------------------------------------------------
Parameter set  Z271* is obtained by adding an isovector-vector nonlinear term 
with the coupling constant $\Lambda_V$ in the Lagrangian density of  Z271 parameter 
set and followed by an adjustment of the $ g_{\rho}$ and 
$\Lambda_V$~\cite{Horowitz01}. On the other hand the  G2* parameter set is
obtained by using a similar procedure. However, since  
the ERMF model  already contains an isovector-vector nonlinear term, the 
G2* parameter set is obtained by merely an adjustment of the  $g_{\rho}$ and  
$\eta_{\rho}$ parameters, keeping the symmetry energy at 
the same value with the G2 parameter set, i.e., $E_{\rm sym}= 24.1$ MeV at 
$k_F =1.14$ fm$^{-1}$~\cite{anto06}. 
%----------------------------------------------------------------------------

By comparing the low density instability regions 
for matter with and without neutrino trapping obtained from both parameter sets we can investigate the role 
of isovector terms and the correlation between the instability regions  
and the symmetry energy. Furthermore, in this study we use the zero 
temperature approximation. We note that 
in the real situation the temperature of protoneutron stars 
is not equal to zero and a supernovae inner core can have a 
temperature around $T \sim$ (10--50) MeV. Indeed, the stability of uniform matter
is sensitive to temperature. This indicates that investigations at 
finite temperature will need to be addressed in the future.

In our approximation, the following constraints can be  used 
to determine the fraction of every constituent in matter which are
later used to calculate the Fermi momentum of every constituent involved:
\begin{itemize}
\item the balance equation for the chemical potentials
\be
\mu_{n}+\mu_{\nu_e} = \mu_{p}+\mu_{e} ~,
\ee
\item conservation of the charge neutrality
\be
\rho_e =\rho_p ~,
\ee
\item and fixed electronic-leptonic fraction  
\be
Y_{l_e}=Y_e+Y_{\nu_e}~,
\ee
\end{itemize}
where the total baryon density is limited by
\be
\rho_B =\rho_n+\rho_p ~.
\ee
Note that in the case of matter without neutrino trapping we have 
$Y_{\nu_e}=0$ and  the value of $Y_{l_e}$ is not fixed. 

\begin{figure*}
\mbox{\epsfig{file=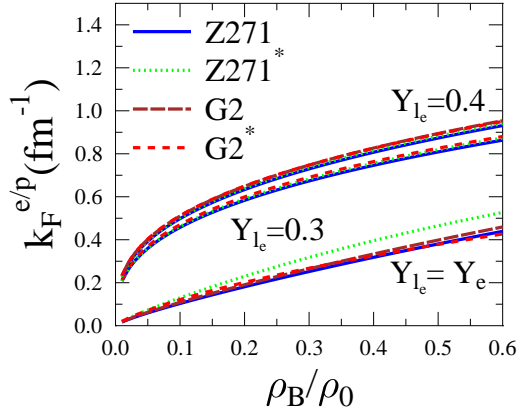,width=8cm}}
\caption{(Color online) Proton (electron) Fermi momentum  
  as a function of the ratio between baryon and nuclear saturation 
  densities for the G2, G2*, Z271, and Z271* parameter sets in the 
neutrinoless matter and in matter with neutrino trapping with $Y_{l_e}=
0.3$ and 0.4.\label{fig:kfermi}}
\end{figure*}

As has been reported in the previous work~\cite{anto06}, besides
the EOS, at low density regimes the asymmetry between the proton and
neutron number ($\alpha$=$Y_N-Y_P$) in matter with and without neutrino
trapping (NT) is also different, i.e., $\alpha^{\rm without ~NT}$ is closer to
the asymmetry of the pure neutron matter (PNM), whereas  
$\alpha^{\rm with ~NT}$ is closer to 
the symmetric nuclear matter (SNM). Thus, $\alpha^{\rm without ~NT}$
has a strong correlation with $a_{\rm sym}$, in contrast
to the $\alpha^{\rm with ~NT}$. This behavior is also found 
in the properties of Fermi 
momentum of each constituent in matter. 
Another different phenomenon is that in
matter with NT we have $k_F^e=k_F^p\sim k_F^n$, 
while matter without NT has $k_F^e= k_F^p\ll k_F^n$. The latter 
indicates that the role of isovector contribution is more significant 
in matter without NT than in matter with NT, 
while the Coulomb interaction has a 
more significant effect in matter with NT due to the presence of more
protons and electrons. Note that Fermi momentum of
every constituent is one of the required information,
besides the nucleon effective mass, to calculate
the polarizations in the longitudinal dielectric function. 
In Fig.~\ref{fig:kfermi}, the effects of the neutrinos in
matter on the electron or proton Fermi momentum at low density regimes
are shown. 

\begin{figure*}
\mbox{\epsfig{file=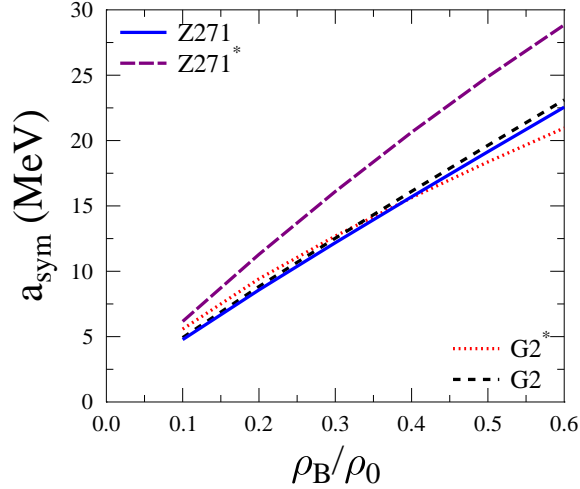,width=8cm}}
\caption{(Color online) Symmetry energy as a function of 
  the ratio between baryon and nuclear saturation 
  densities for the G2, G2*, Z271, and Z271* parameter sets.\label{fig:asym}}
\end{figure*}

The symmetry energies 
($a_{\rm sym}$) of the corresponding parameter sets are shown in
Fig.~\ref{fig:asym}. It is clearly seen that, different from the
Horowitz-Piekarewicz model, the  high
density adjustment in the isovector-vector channel of the Furnstahl-Serot-Tang
model does not significantly affect its $a_{\rm sym}$ at low density. On the
other hand, parameter set with softer symmetry energy at high density
of Horowitz-Piekarewicz model becomes stiffer at low density
regimes. It means that, in contrast to the Furnstahl-Serot-Tang model, the
high density isovector adjustment in the Horowitz-Piekarewicz model leads
to a more repulsive isovector interaction than the standard one at low
density regimes. The different forms of the nonlinear terms in both 
models are responsible for these different behaviors. 

From both figures, it can be explicitly seen that the presence 
of neutrinos in matter leads to higher Fermi
momenta  of protons and electrons compared to the case 
of  matter without NT, while in the latter 
Fermi momentum of every constituent, which is represented by protons and
electrons, is independent to the model used. Furthermore,
in matter without neutrino trapping, due to its proton-neutron
asymmetry that is closer to PNM, the Fermi momentum of 
each constituent can be correlated to the $a_{\rm sym}$.

%%%%%%%%%%%%%%%%%%%%%%%%%%%%%%%%%%%%%%%%%%%%%%%%%%%%%%%%%%%%%%%%%%%%%%%%%
\section{Longitudinal Dielectric Function}
\label{sec_LDF}
%%%%%%%%%%%%%%%%%%%%%%%%%%%%%%%%%%%%%%%%%%%%%%%%%%%%%%%%%%%%%%%%%%%%%%%%%
The longitudinal dielectric function can be written as~\cite{Carr}
\be
\varepsilon_L (q,q_0)= {\rm{det}} \[1-D_L(q,q_0) \Pi_L(q,q_0)\] . \label{eq:det}
\ee 
The uniform ground state system becomes unstable to small-amplitude
density fluctuations with perturbation momentum $q$ when $\varepsilon_L (q,q_0=0)=\varepsilon_L \le 0$. 
Note that in Eq.~(\ref{eq:det}) $q_0$ is the time-component of the
four-momentum  $q^\mu=(q_0,\vec{q}\,)$ and $q=|\vec{q}\,|$.
The critical density $\rho_c$ is the largest
density for which the above condition has a solution.
For matter consisting of protons, neutrons, and electrons, the 
longitudinal meson propagator is given by
\bea
D_L = \left( \begin{array} {cccc}
d_g& 0 & -d_g & 0\\
0& -d_s & d_{sv\rho}^+ & d_{sv\rho}^-\\
-d_g &  d_{sv\rho}^+ & d_{33} & d_{v\rho}^-\\
0& d_{sv\rho}^- &  d_{v\rho}^- & d_{44}
\end{array}\right) ~,
\label{eq:longprop}
\eea
where $d_{sv\rho}^+ = - (d_{sv} +   d_{s\rho}$), $d_{sv\rho}^- =
-(d_{sv} -   d_{s\rho}$), $d_{v\rho}^- =  d_{v} -   d_{\rho}$, $d_{33}
= d_g +  d_{v} + d_{\rho} + 2  d_{v\rho}$ and $d_{44} = d_{v} +
d_{\rho} - 2  d_{v\rho}$. In this form, mixing propagators between
isoscalar-scalar and isoscalar-vector ($d_{sv}$), isoscalar-vector and
isovector-vector ($d_{v\rho}$), isoscalar-scalar and isovector-vector
($d_{s\rho}$) are present due to the nonlinear mixing terms in the model, in addition to the standard  photon, omega, sigma and rho propagators ($d_g$,  $d_{v}$, $d_{s}$ and  $d_{\rho}$). These propagators are determined from the quadratic fluctuations around the static solutions which are generated by the second derivatives of energy density (${\partial^2 \epsilon}/{\partial \phi_i \partial\phi_j}$), where $\phi_i$ and $\phi_j$ are the  involved meson fields.
The explicit forms of the $\sigma$, $\omega$, and $\rho$ propagators are
\bea
d_s&=&\frac{g_{\sigma}^2 (q^2+m_\omega^{*~2})(q^2+m_\rho^{*~2})}{(q^2+m_\omega^{*~2})(q^2+m_\rho^{*~2})(q^2+m_\sigma^{*~2})+(\Pi_{\sigma \omega}^0)^2(q^2+m_\rho^{*~2})+(\Pi_{\sigma \rho}^0)^2 (q^2+m_\omega^{*~2})}~,\\
d_v&=&\frac{g_{\omega}^2 (q^2+m_\sigma^{*~2})(q^2+m_\rho^{*~2})}{(q^2+m_\omega^{*~2})(q^2+m_\rho^{*~2})(q^2+m_\sigma^{*~2})+(\Pi_{\sigma \omega}^0)^2(q^2+m_\rho^{*~2})-(\Pi_{\omega \rho}^{00})^2 (q^2+m_\sigma^{*~2})}~,\\
d_\rho&=&\frac{1/4 g_\rho^2 (q^2+m_\sigma^{*~2})(q^2+m_\omega^{*~2})}{(q^2+m_\omega^{*~2})(q^2+m_\rho^{*~2})(q^2+m_\sigma^{*~2})+(\Pi_{\sigma \rho}^0)^2(q^2+m_\omega^{*~2})-(\Pi_{\omega \rho}^{00})^2 (q^2+m_\sigma^{*~2})}~,
\label{eq:d3}
\eea
and in the mixing propagators,
\bea
d_{sv}&=&\frac{g_{\sigma} g_{\omega} \Pi_{\omega \sigma}^{0}(q^2+m_\rho^{*~2})}{H(q,q_0=0)}~,\\
d_{s\rho}&=&\frac{1/2 g_\rho g_{\sigma} \Pi_{\sigma \rho}^{0}(q^2+m_\omega^{*~2})}{H(q,q_0=0)}~,\\
d_{v \rho}&=&\frac{1/2 g_\rho g_{\omega} \Pi_{\omega \rho}^{00}(q^2+m_\sigma^{*~2})}{H(q,q_0=0)}~,
\eea
with
\bea
H(q,q_0=0)&=&(q^2+m_\omega^{*~2})(q^2+m_\rho^{*~2})(q^2+m_\sigma^{*~2})+(\Pi_{\sigma \omega}^0)^2(q^2+m_\rho^{*~2})\nonumber\\&+&(\Pi_{\sigma \rho}^0)^2 (q^2+m_\omega^{*~2})-(\Pi_{\omega \rho}^{00})^2 (q^2+m_\sigma^{*~2})~,
\eea
where the effective mass of each meson is
\bea
m_\sigma^{*~2}&=& \frac{\partial^2 \epsilon}{\partial^2 \sigma}=m_\sigma^2+2 b_2 \sigma+3 b_3 \sigma^2 - d_3  V_0^2 -  \tilde{\Lambda}_s  b_0^2~,\\ m_\omega^{*~2}&=& -\frac{\partial^2 \epsilon}{\partial^2 V_0}=m_\omega^2+2 d_2\sigma+d_3 \sigma^2 + 3 c_3  V_0^2 +  \tilde{\Lambda}_v  b_0^2~, \\m_\rho^{*~2}&=& -\frac{\partial^2 \epsilon}{\partial^2 b_0}=m_\rho^2+2 f_2\sigma + \tilde{\Lambda}_s \sigma^2 +  \tilde{\Lambda}_v V_0^2~,
\label{eq:meseffmass}
\eea
and the effective mixing masses read
\bea
\Pi_{\sigma \omega}^0&=&-\frac{\partial^2 \epsilon}{\partial \sigma \partial V_0}= 2 d_2 V_0 + 2 d_3 \sigma V_0~,\\ \Pi_{\sigma \rho}^0&=&-\frac{\partial^2 \epsilon}{\partial \sigma \partial b_0}= 2 f_2 b_0 + 2 \tilde{\Lambda}_s \sigma b_0~,\\ \Pi_{\omega \rho}^{0 0}&=&\frac{\partial^2 \epsilon}{\partial V_0 \partial b_0}= -2 \tilde{\Lambda}_v V_0 b_0 ~,
\label{eq:polar}
\eea
whereas the propagator of photon (Coulomb) is 
\bea
d_g = \frac{e^2}{q^2} ~.
\eea
The longitudinal polarization matrix given in Eq.~(\ref{eq:det}) reads
\bea 
\Pi_L = \left( \begin{array} {cccc}
\Pi_{00}^e & 0 & 0&0\\
0& \Pi_{s}& \Pi_{m}^p &\Pi_{m}^n\\
0& \Pi_{m}^p & \Pi_{00}^p & 0\\
0& \Pi_{m}^n & 0 & \Pi_{00}^n
\end{array}\right) ~.
\label{eq:longpol}
\eea

\begin{figure*}
\mbox{\epsfig{file=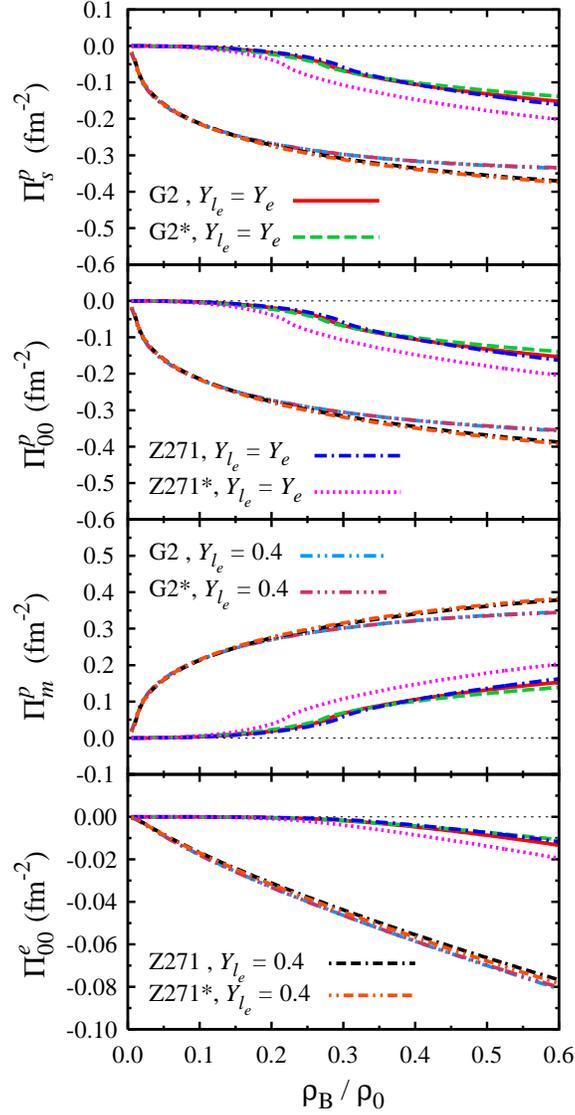,width=9cm}}
\caption{(Color online) The proton-scalar, -longitudinal, 
-mixing, and electron-longitudinal
polarizations as a function of the ratio between baryon and nuclear saturation 
  densities for the G2, G2*, Z271, and Z271* parameter sets in
neutrinoless matter and matter with neutrino trapping with $Y_{l_e}=
0.4$. All curves have been obtained by using 
$q = 0.5$ $\rm fm^{-1}$.\label{fig:polar}}
\end{figure*}

\begin{figure*}
\mbox{\epsfig{file=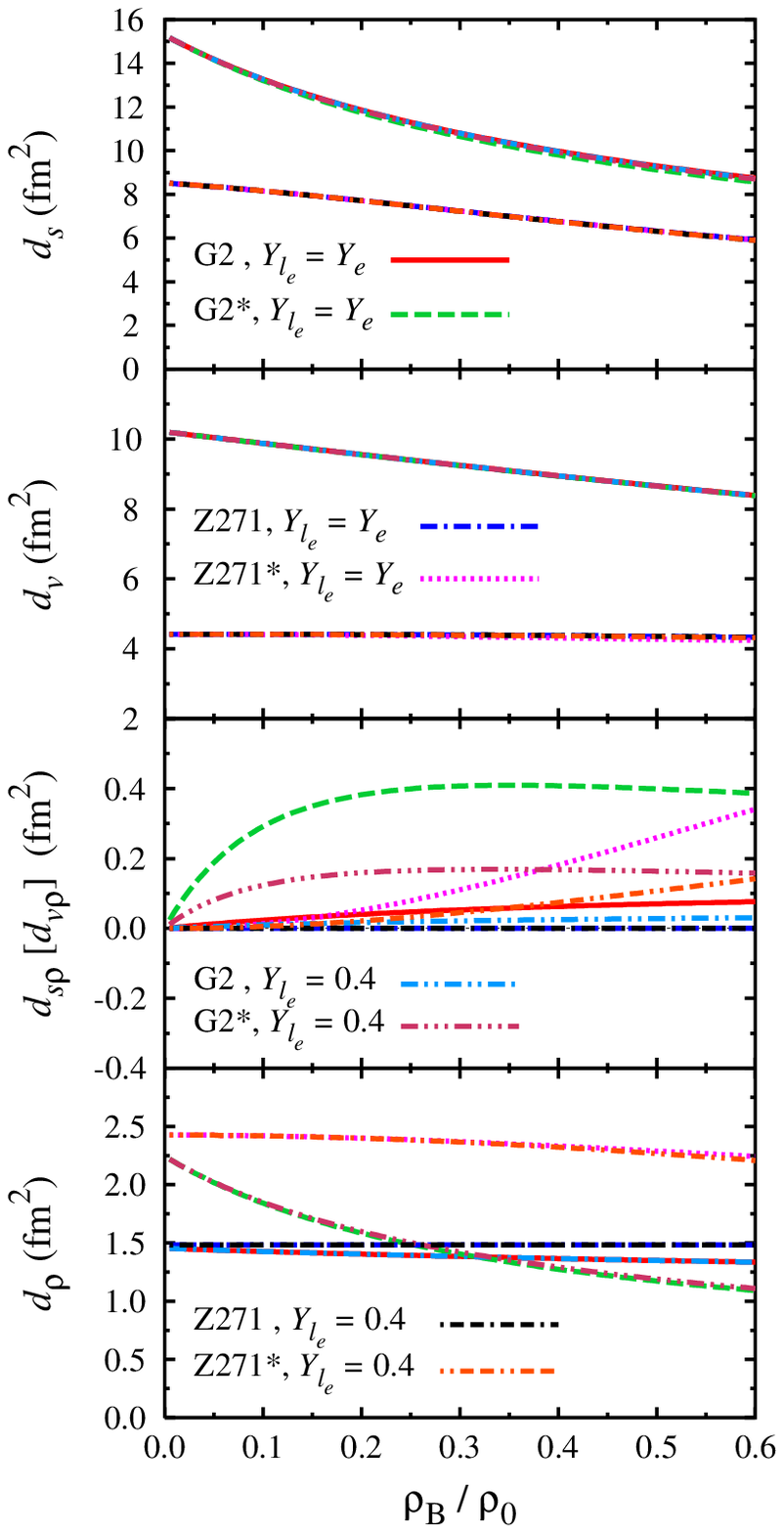,width=9cm}}
\caption{(Color online) The scalar, vector, rho and 
  mixing scalar-rho or vector-rho propagators as a function of the ratio between baryon and nuclear saturation 
  densities for the G2, G2*, Z271, and Z271* parameter sets in
neutrinoless matter and matter with neutrino trapping with $Y_{l_e}=
0.4$ and  $q = 0.5$ $\rm fm^{-1}$.
%-------------------------------------------------------------
Note that for the Horowitz-Piekarewicz model $d_{s\rho}$ = 0,
whereas for the Furnstahl-Serot-Tang  model $d_{v\rho}$=0.
%-------------------------------------------------------------
\label{fig:propa}}
\end{figure*}

The formulas for polarization elements in $\Pi_L$ are given in, e.g.,
Ref.~\cite{anto06}. Note that for the Horowitz-Piekarewicz model
$\Pi_{\sigma \omega}^0$ and $\Pi_{\sigma \rho}^0$ equal to
zero. On the other hand, for the Furnstahl-Serot-Tang model, $\Pi_{\omega
\rho}^0$ and 
$\tilde{\Lambda}_s$ in  $\Pi_{\sigma \rho}^0$ are zero. In
Figs.~\ref{fig:polar} and~\ref{fig:propa}, we show some plots 
as examples of the  polarizations
and propagators at $q = 0.5$ $\rm fm^{-1}$. It is clearly seen that all
polarizations depend strongly on whether the neutrinos are trapped or
not in matter. In matter with NT, at a density close to 0.6
$\rho_B$, they have only a weak dependence on the model used 
due to the increasing role of the nucleon
effective mass. For matter without NT, a similar model dependence
on its Fermi momentum $k_F$ shows up (see Fig.~\ref{fig:kfermi}). 
This indicates that  the polarizations can be correlated 
with $a_{\rm sym}$. Note that the nucleon effective mass has
no correlation with the symmetry energy 
$a_{\rm sym}$~\cite{anto06}. This demonstrates that, instead
of the effective mass of nucleons, Fermi momentum of each constituent
controls the behavior of each polarization.

Figure~\ref{fig:propa} shows the behavior of the propagators as a function
of the ratio between baryon and nuclear saturation densities. 
The scalar ($d_s$), vector  ($d_v$) and rho ($d_\rho$) propagators 
are clearly model dependent and they are insensitive to the
presence of the neutrino (NT) in matter. Only  $d_\rho$ depends on the 
isovector adjustment. The presence of neutrinos has a significant effect 
only in the mixing propagators ($d_{s \rho}$ and $d_{v\rho}$). However,
their contributions are smaller compared to the rest. We also note that 
the propagator $d_{s v}$, which appears in the Furnstahl-Serot-Tang 
model, has an order of magnitude that equals to the propagators $d_{s \rho}$ 
and $d_{v\rho}$, and, furthermore, it does not depend on the adjustment
of the isovector sector. These properties are a manifestation of the 
interplay among $\sigma$, $\omega$ and $\rho$ meson effective masses as well
as mixing effective masses $\Pi_{\sigma \omega}^0$,
$\Pi_{\sigma \rho}^0$ and $\Pi_{\rho \omega}^0$.

To simplify our discussion, let us neglect the minor contributions of the 
mixing propagators. In this picture we obtain $d^{\rm with~NT}\simeq d^{\rm
without~NT}$. Together with the fact that $d_\rho$ depends on the adjustment of 
isovector sector (model dependent), this 
indicates that $d_\rho$ can be related with $a_{\rm sym}$. 

Therefore, in the case of matter with NT, we can understand  that the
critical density is higher and the instability region is larger 
than those in the case without NT. Furthermore, the fact that 
both of them are insensitive to the value of $Y_{l_e}$ and 
they are not influenced by the isovector treatment appears as a
consequence of the small proton-neutron asymmetry $\alpha$, 
which makes the role of the polarizations more dominant compared to 
the role of 
the corresponding propagators, in controlling the behavior of the 
longitudinal dielectric function. The correlation between critical
density, as well as the onset of the instability, and $a_{\rm sym}$ 
in matter without NT are caused by two sources: the dependence of 
the polarization on the $a_{\rm sym}$ as well as on the isovector propagators 
($d_{\rho}$, $d_{s \rho}$ and $d_{v\rho}$). 
%%%%%%%%%%%%%%%%%%%%%%%%%%%%%%%%%%%%%%%%%%%%%%%%%%%%%%%%%%%%%%%%%%%%%%%%%
\section{The Onset of Instability}
\label{sec_results}
%%%%%%%%%%%%%%%%%%%%%%%%%%%%%%%%%%%%%%%%%%%%%%%%%%%%%%%%%%%%%%%%%%%%%%%%%

\begin{figure*}
\centering
 \mbox{\epsfig{file=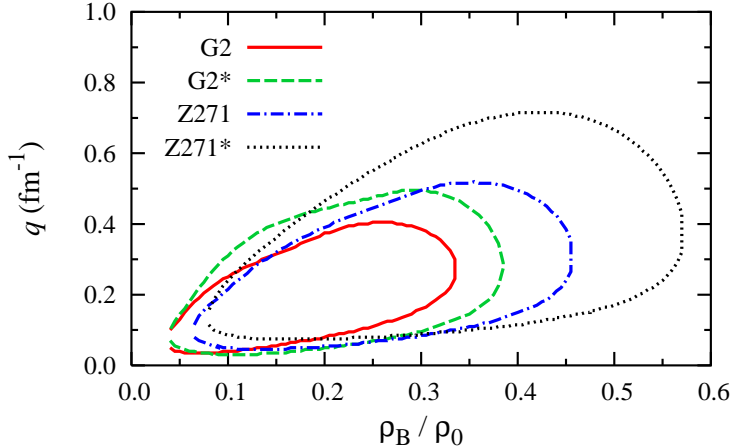,width=10cm}}
\caption{(Color online) Onset of the instability of 
  the neutrinoless matter  obtained by using the Horowitz-Piekarewicz 
%-----------------------------------------------------------------------
(Z271 and Z271*) and  Furnstahl-Serot-Tang (G2 and G2*) models
%---------------------------------------------------------------------
  as functions of the ratio between baryon 
  and nuclear saturation densities and the perturbation momentum~$q$.  
  \label{fig:h-p-without}}
\end{figure*}

As mentioned in the Introduction, here we intend to investigate every
point in the onset
of the instability region in a great detail. To this end, we plot the
projection of the longitudinal dielectric function given in 
Eq.~(\ref{eq:det}) on the 
$\rho/\rho_0-q$ plane in the case that
$\varepsilon_L =0$. To obtain more information on the role 
of the Coulomb interaction in the limit of $q \approx 0$ (almost
zero perturbation), we also present the longitudinal dielectric 
function at $q$ close to zero, i.e.,
$\varepsilon_L (q=0.01 ~{\rm fm^{-1}},q_0=0)$, as 
a function of the $\rho/\rho_0$. 

\begin{figure*}
\centering
 \mbox{\epsfig{file=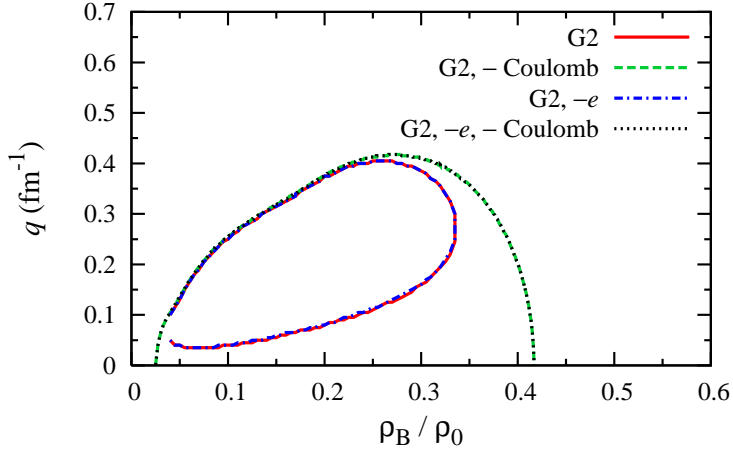,width=10cm}}
\caption{(Color online) Onset of the instability of the 
 neutrinoless matter obtained by using the 
 Furnstahl-Serot-Tang model with and without (symbolized with $-$) Coulomb
 interaction and electrons contributions. 
 Note that the solid (red) and dash-dotted (blue) curves [as well as the dashed (green) 
 and dotted (black) ones]
 are coincident.}
 \label{fig:fst-leptons}
\end{figure*}

\begin{figure*}
\centering
 \mbox{\epsfig{file=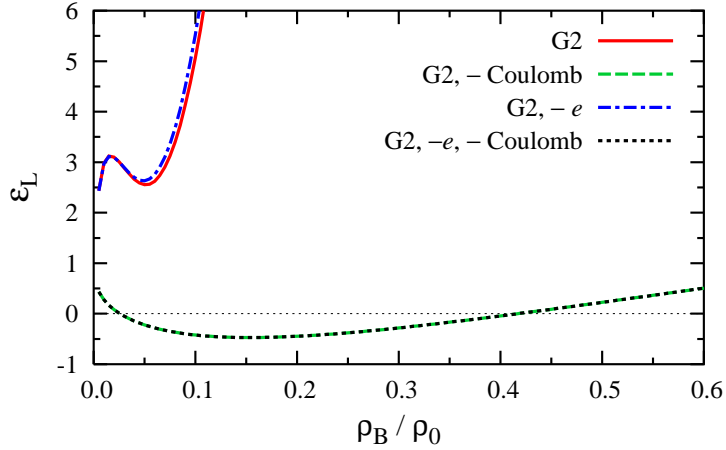,width=10cm}}
\caption{(Color online) Longitudinal dielectric function for
 matter without NT obtained by using the 
 Furnstahl-Serot-Tang model with and without (symbolized with $-$) Coulomb
 interaction as well as 
 electrons contributions. Note that the dashed (green) and dotted (black) curves 
 are coincident. 
 All curves have been calculated by using $q=0.01~{\rm fm^{-1}}$.}
 \label{fig:fst-leptons-a}
\end{figure*}

The effects of the isovector-vector channel adjustment in both models
are exhibited in Fig.~\ref{fig:h-p-without}.
Obviously, from the size and the position of the boundary of the instability region, the 
adjustment in the isovector-vector sector has a more significant effect 
in the Horowitz-Piekarewicz model compared with the Furnstahl-Serot-Tang one.
It is also clear from the figure that this adjustment leads to a higher 
critical density in both models. This result is certainly consistent 
with our previous study (cf. Fig. 8 of Ref.~\cite{anto06} in the case
of $Y_{l_e}=Y_e$). Another important 
finding obtained from these figures is that in both cases the ERMF model of 
Furnstahl-Serot-Tang yields a smaller onset of instability.  
%-----------------------------------------------------------------------
This is due to the fact that the parameter sets Z271* and G2* 
have larger symmetry energies compared to the Z271 and G2. 
Therefore, beta-equilibrium of these parameter sets
is attained with larger proton fractions and a smaller
contribution from the repulsive isovector channel.
%-------------------------------------------------------------------------
Thus, as we expected, Fig.~\ref{fig:h-p-without} shows explicitly the correlation between the 
onset of the instability region with  $a_{\rm sym}$. 
The reason of this fact has been explained in the previous section.

In Fig.~\ref{fig:fst-leptons}, 
we show the effect of the electron absence (indicated by ``$-e$'' in the figure) on 
the onset of instability
by switching off their contribution in the case of 
matter without NT using G2 parameter set, with and without Coulomb 
contribution. Note that the latter is indicated with ``$-$Coulomb'' in the figure. 
The effect of the electron presence on the size of the instability 
region is found to be negligible in both cases. The reason is that 
the number of the electrons is too small in matter without NT, and therefore, 
the effect of the attractive Coulomb interaction generated by electrons and protons 
is too weak to produce a visible impact on every point in the onset of
instability.  Such behavior is observed 
even in the limit of $q$ close to zero (see Fig.~\ref{fig:fst-leptons-a}).

On the other hand, the repulsive Coulomb interaction due to the
presence of protons, even in a very small number, enlarges moderately the
stability region of this matter in the range of 
$0.05\le\rho/\rho_0$ $\le$ 0.4, which is clearly shown
in Fig.~\ref{fig:fst-leptons}. Furthermore, Fig.~\ref{fig:fst-leptons-a} 
emphasizes and shows the
important role of Coulomb interaction to stabilize matter without
NT for almost zero perturbation.

%%%%%%%%%%%%%%%%%%%%%%%%%%%%%%%%%%%%%%%%%%%%%%%%%%%%%%%%%%%%%%%%%%%%%%%%%
\begin{figure*}
\centering
 \mbox{\epsfig{file=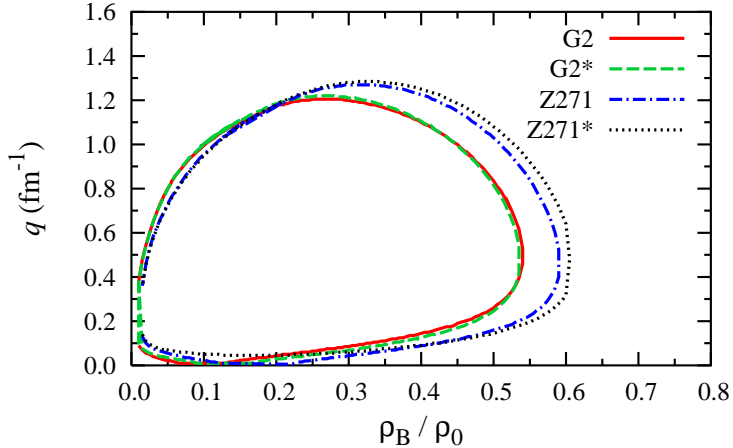,width=10cm}}
\caption{(Color online) Onset of the instability for matter with NT as 
  functions of the ratio between baryon and nuclear saturation densities 
  and the perturbation momentum $q$. All curves have been obtained by using
  $Y_{l_e}=0.3$.
%-----------------------------------------------------------------------
 The results are obtained by using the Horowitz-Piekarewicz 
(Z271 and Z271*) and Furnstahl-Serot-Tang (G2 and G2*) models.
 %----------------------------------------------------------------------- 
  \label{fig:h-p-with1}}
\end{figure*}

\begin{figure*}
\centering
 \mbox{\epsfig{file=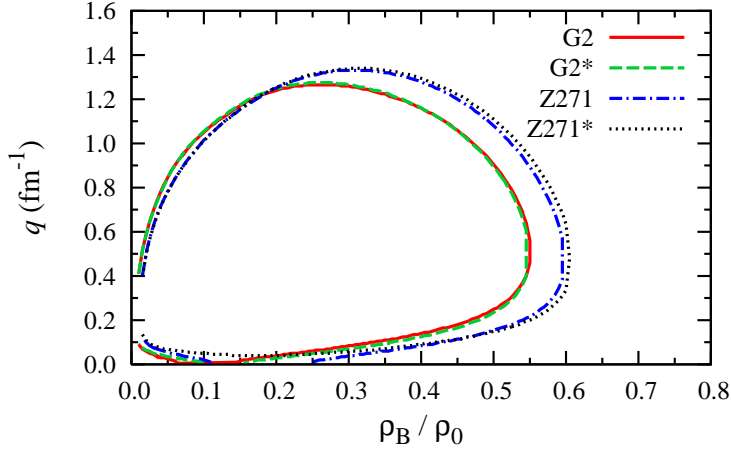,width=10cm}}
\caption{(Color online) As in Fig.~\ref{fig:h-p-with1}, but for 
  $Y_{l_e}=0.4$.
  \label{fig:h-p-with2}}
\end{figure*}

When the neutrino contribution is taken into account, the situation 
dramatically changes. This is shown in Fig.~\ref{fig:h-p-with1} 
for $Y_{l_e}$= $0.3$ and in Fig.~\ref{fig:h-p-with2} for 
$Y_{l_e}$= $0.4$.   For all parameter sets the instability
boundaries expand, for which no substantial difference appears 
in the onset of the instability due to the isovector adjustment, 
except for the region with $q$ close to zero. In this region, the 
variation of the neutrino fraction in matter also yields an 
insignificant effect in the onset of
instability, as we have expected. 

\begin{figure*}
\centering
 \mbox{\epsfig{file=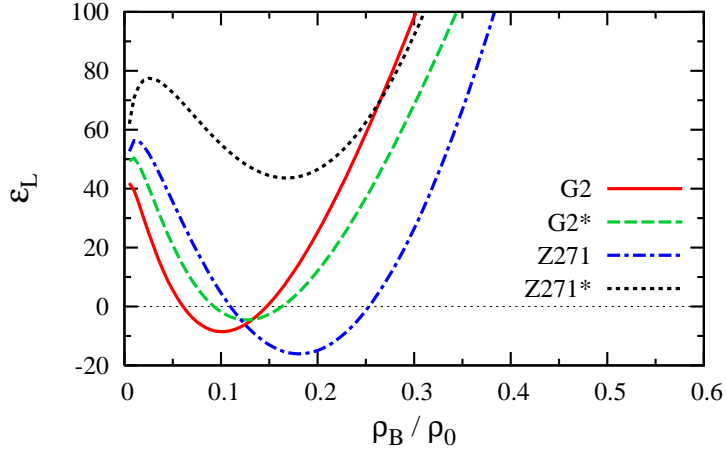,width=10cm}}
\caption{(Color online)  Longitudinal dielectric function for 
  matter with NT as a function of the ratio between baryon 
  and nuclear saturation densities. All curves have been obtained
  by using the perturbation momentum $q =0.01~{\rm fm}^{-1}$
  and $Y_{l_e}=0.4$. 
%-----------------------------------------------------------------------
 The results are obtained by using the Horowitz-Piekarewicz 
(Z271 and Z271*) and Furnstahl-Serot-Tang (G2 and G2*) models.
 %-----------------------------------------------------------------------  
  \label{fig:h-p-with3}}
\end{figure*}

If we observe the longitudinal dielectric function at $q$ close to zero
($q=0.01~{\rm fm}^{-1}$), as shown in
Fig.~\ref{fig:h-p-with3}, then we can clearly see that the Z271* parameter
set shows a quite different behavior compared to the other parameter sets,
i.e., the Z271* parameter always yields $\varepsilon_L >$ 0 in this limit. 
This fact indicates that the transition to more stable
region at the points with  small perturbation $q <$ 0.1  $\rm fm^{-1}$
 around $0.1 \le \rho/\rho_0 \le 0.3$ of the Z271* is driven by a different 
mechanism compared to other parameter sets, i.e.,
%----------------------------------------------------------------------------
it has a larger $a_{\rm sym}$ that leads to larger proton and electron 
fractions (and less neutrinos for a fixed lepton fraction).
As a consequence, a larger repulsion effect is produced compared 
to other parameter sets.

%------------------------------------------------------------------------------
From the fact that the
adjusted parameter sets (indicated with *) 
have narrower instability region compared to
their counterparts (indicated without *) for both models, 
then we can conclude that 
in this limit the onset of the instability is strongly related 
to the adjustment of the isovector sector. 

\begin{figure*}
\centering
 \mbox{\epsfig{file=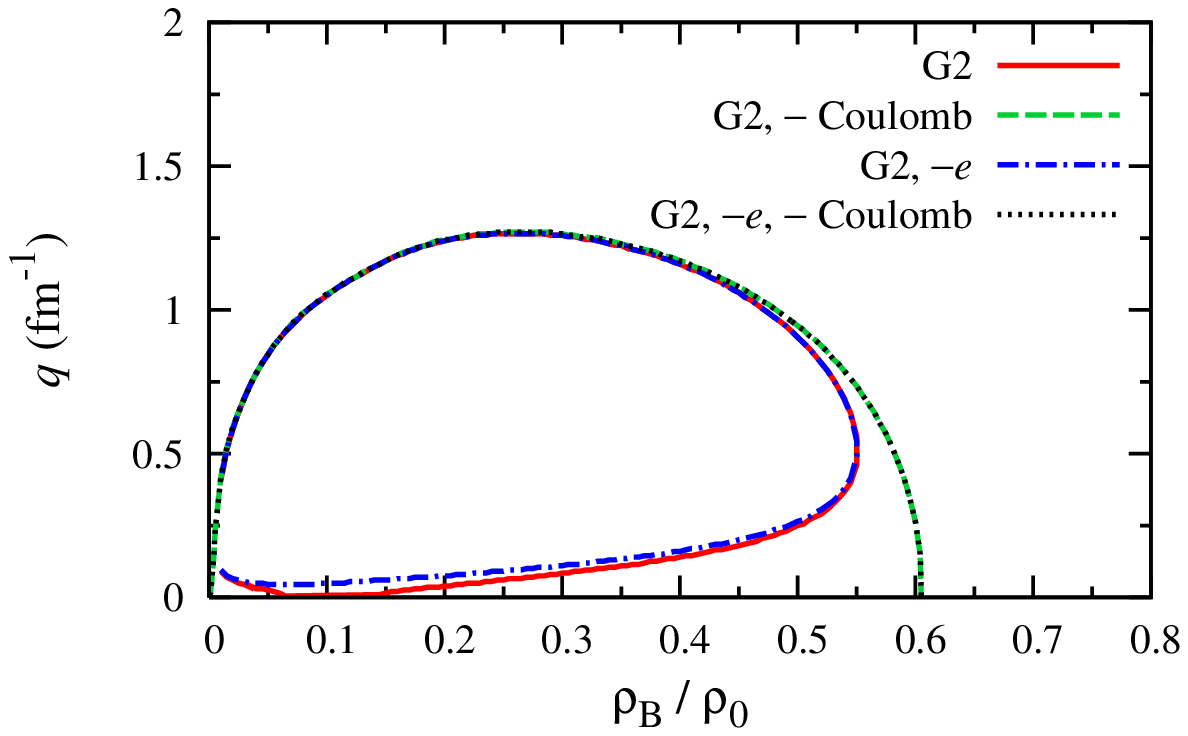,width=10cm}}
\caption{(Color online) As in Fig.~\ref{fig:fst-leptons}, but calculated for
 the case of neutrino trapping with $Y_{l_e}=0.4$.
  \label{fig:h-p-with4}}
\end{figure*}

\begin{figure*}
\centering
 \mbox{\epsfig{file=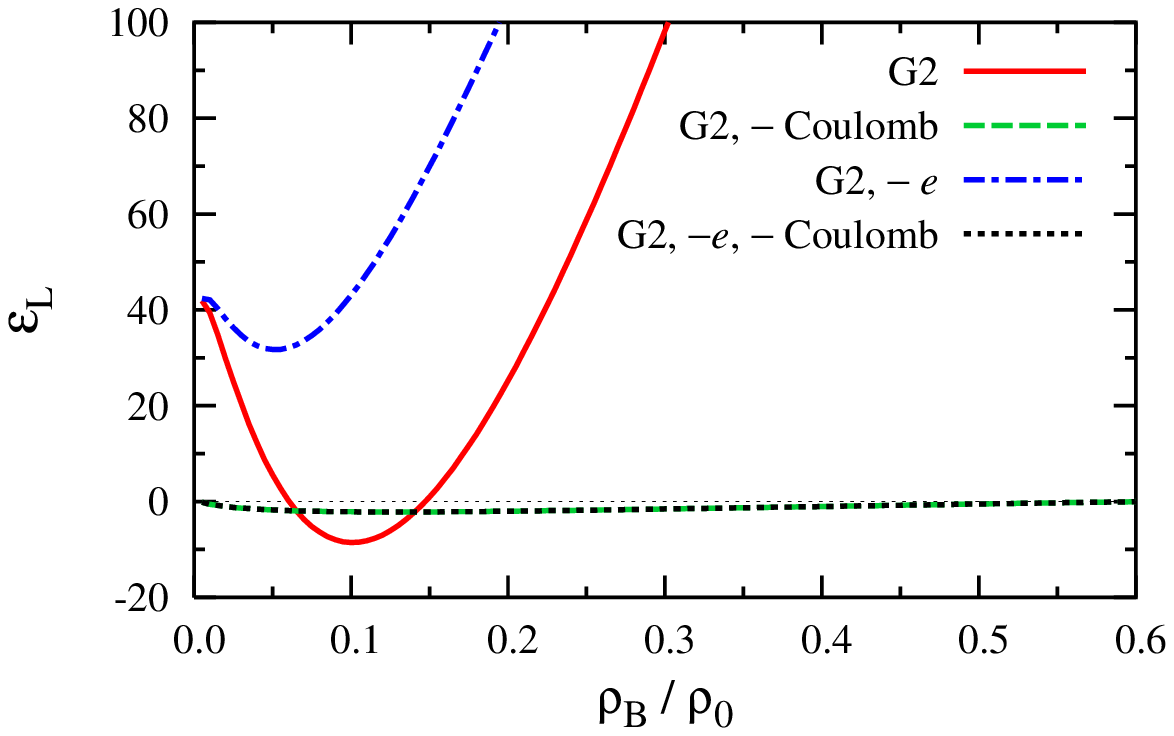,width=10cm}}
\caption{(Color online) As in Fig.~\ref{fig:fst-leptons-a}, but calculated for
  the case of neutrino trapping with $Y_{l_e}=0.4$.
  \label{fig:h-p-with5}}
\end{figure*}

To dig up more information behind this fact, we show in
Figs.~\ref{fig:h-p-with4} and  ~\ref{fig:h-p-with5} the effects 
of the electrons absence (indicated by ``$-e$'' in the figure) 
and the absence of the 
Coulomb interaction (indicated by ``$-$Coulomb) 
on the onset of the instability and on the instability in the limit of $q$
close to zero, by switching off their contributions in the case of 
matter with NT. It can be seen from Fig.~\ref{fig:h-p-with4} that
the Coulomb interaction plays an important role in stabilizing 
the region with $q <$ $0.6$  $\rm fm^{-1}$, in the range of 
$0.2 \le \rho/\rho_0 \le 0.6$. Furthermore, from Fig.~\ref{fig:h-p-with5} 
we can see that if the electrons contribution were turned off, then 
the instability at $q$ close to zero would 
disappear. Thus, for matter with NT the appearance 
of the large and negative $\varepsilon_L$ for the G2, G2* and Z271 parameter sets 
is caused by the fact that 
the repulsive interaction induced by the 
proton-neutron asymmetry (isovector) is unable to cancel 
the strong-attractive Coulomb interaction created by 
the presence of a substantially large number of electrons in matter.

Figure~\ref{fig:h-p-with5} also shows that if the electrons 
are present but their Coulomb interactions were turned off
(electrons behave as free particles) then the longitudinal
dielectric function $\varepsilon_L$ became
larger and closer to $\varepsilon_L =0$ but the instability
boundary enlarges. This result can be used to emphasize the important
role of the Coulomb interaction to stabilize the matter with NT in the
limit of $q$ close to zero.

%%%%%%%%%%%%%%%%%%%%%%%%%%%%%%%%%%%%%%%%%%%%%%%%%%%%%%%%%%%%%%%%%%%%%%%%%
\section{Conclusion}
\label{sec_conclu}
%%%%%%%%%%%%%%%%%%%%%%%%%%%%%%%%%%%%%%%%%%%%%%%%%%%%%%%%%%%%%%%%%%%%%%%%%
We have studied how the instability region starts to appear in low-density matter
described by the Horowitz-Piekarewicz and Furnstahl-Serot-Tang models. 
To this end we have utilized the longitudinal dielectric function at $q_0=0$.
The importance of the electron and Coulomb terms in matter with  
neutrino trapping has been investigated. It is found that the adjustment
of the isovector terms has a more significant effect in
the Horowitz-Piekarewicz model, i.e., producing a stronger repulsive isovector
contribution which leads to a stronger correlation between
its low density instability region and the $a_{\rm sym}$ compared
to the model of Furnstahl-Serot-Tang for matter without neutrino
trapping. 
%----------------------------------------------------------------------------
In the case of matter with neutrino trapping, the parameter 
sets with stiff EOS at low density lead to
a large and positive $\varepsilon_L (q,q_0=0)$. This demonstrates that, 
although the onsets of the instability of 
parameter sets with stiff and soft EOS at low densities are similar, the
driving mechanisms are different. 
%----------------------------------------------------------------------------
This fact might have an effect to 
the neutrino transport in matter.  
In both models the effect of the variation of the leptonic fraction is negligible,
but the effect of the neutrino trapping on the onset of 
the instability region is significant. The presence of more protons and electrons 
in matter with neutrino trapping 
is the reason behind this phenomenon.
The Coulomb term is found to
be decisive in enlarging the stability of matter
in this density region. The presence of the large and 
negative $\varepsilon_L (q,q_0=0)$ in 
some parts of the instability region of matter with neutrino 
trapping originates from 
the fact that the isovector term is insufficient to
cancel the attractive Coulomb interaction contributions generated by 
the presence of electrons (and protons).

%%%%%%%%%%%%%%%%%%%%%%%%%%%%%%%%%%%%%%%%%%%%%%%%%%%%%%%%%%%%%%%%%%%%%%%%%
\section*{ACKNOWLEDGMENT}
Support from the University of Indonesia 
is gratefully acknowledged.
%%%%%%%%%%%%%%%%%%%%%%%%%%%%%%%%%%%%%%%%%%%%%%%%%%%%%%%%%%%%%%%%%%%%%%%%%

\end{document}